# High speed free-space optical communication using standard fiber communication component without optical amplification

Yao Zhang,[1] Hua-Ying Liu,[1, *] Xiaoyi Liu,[1] Peng Xu,[1] Xiang Dong,[1] Pengfei Fan,[1] Xiaohui Tian,[1] Hua Yu,[2] Dong Pan,[3] Zhijun Yin,[2] Guilu Long,[3] Shi-Ning Zhu,[1] and Zhenda Xie[1, *]

[1]*National Laboratory of Solid State Microstructures, School of Electronic Science and Engineering, School of Physics, College of Engineering and Applied Sciences, and Collaborative Innovation Center of Advanced Microstructures, Nanjing University, Nanjing 210093, China*
[2]*Nanzhi Photonlink Technology co., Ltd., Ningbo 315000，China*
[3]*Beijing Academy of Quantum Information Sciences, Beijing 100193, China*
*Corresponding author: liuhuaying@nju.edu.cn and xiezhenda@nju.edu.cn*

**Abstract:** Free-space optical communication (FSO) can achieve fast, secure and license-free communication without need for physical cables, making it a cost-effective, energy-efficient and flexible solution when the fiber connection is unavailable. To establish FSO connection on-demand, it is essential to build portable FSO devices with compact structure and light weight. Here, we develop a miniaturized FSO system and realize 9.16 Gbps FSO between two nodes that is 1 km apart, using a commercial single-mode-fiber-coupled optical transceiver module without optical amplification. Using our 4-stage acquisition, pointing and tracking (APT) systems, the tracking error is within 3 μrad and results an average link loss of 13.7 dB, which is the key for this high-bandwidth FSO demonstration without optical amplification. Our FSO link has been tested up to 4 km, with link loss of 18 dB that is limited by the foggy weather during the test. Longer FSO distances can be expected with better weather condition and optical amplification. With single FSO device weight of only 9.5 kg, this result arouses massive applications of field-deployable high-speed wireless communication.

Free-space optical communication (FSO) has received increasing attention as an alternate to the fiber communication, for high-bandwidth wireless data transmission[1-4], and it can be built in different scales for various applications. In space-wide scale, FSO is the key to establish high-speed satellite internet like Starlink for global network coverage.[5-7] While in the near-ground low-altitude scale, FSO is also attractive for its capability of providing high-data-rate, license-free and high security connection, in a relatively small region, for applications including last-mile connection, disaster recovery and military communication etc.[8-10] In these flexible communication scenarios, the FSO system should be miniaturized, integrated, and lightweight for easy-handling as portable devices. However, it can be challenging to pack acquisition, pointing and tracking (APT) systems in small sizes while keeping high tracking accuracy for sufficiently low link loss, especially over some distance. One common method is to use optical amplifiers like erbium-doped optical fiber amplifier (EDFA)[11-14] to compensate for the high link loss, but that inevitably increase system complexity, power consumption, and weight, from the other extent.

Here we demonstrate FSO at 1 km distance with communication bandwidth of 9.16 Gbps, using a pair of commercial fiber optical communication transceiver modules without optical amplification. Low average link loss of 13.7 dB, for the coupling into single mode fiber (SMF), is measured using our FSO devices, which is key for this demonstration of bidirectional optical communication. Such low link loss is enabled by the low-diffraction optical design, fully automatic 4-stage close-loop feedback control, and the motion stabilization system in our APT units. Without the need of optical amplifiers, a single FSO device only weighs 9.5 kg, within a size of 45×40×35 cm$^3$. It can be used in a pair for a single link or more pairs for multiple links, in a plug-and-play manner to establish FSO channels in minutes, and thus fulfills the demand of field-deployable high-speed data transmission.

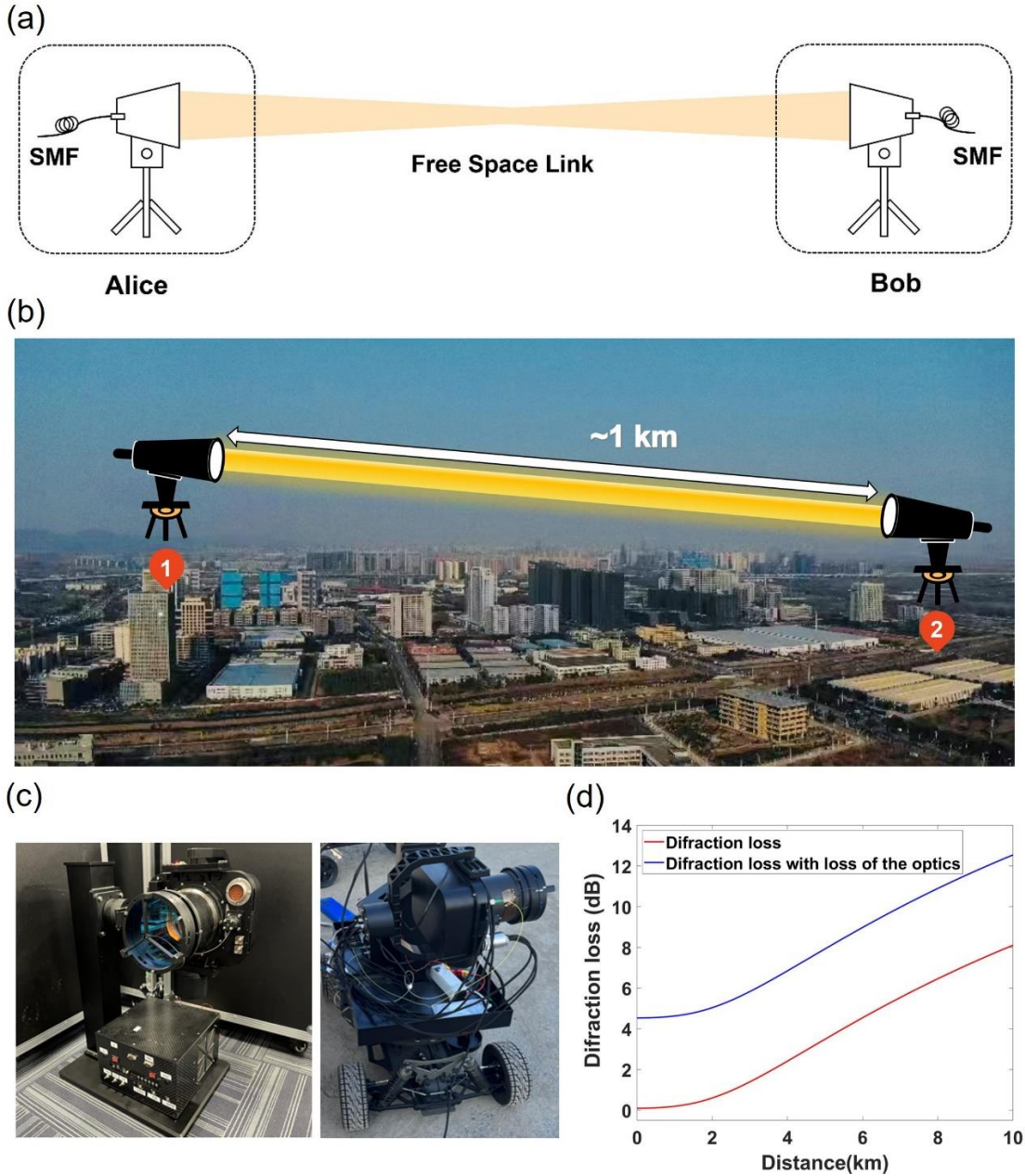

**FIG. 1.** (a) Scheme of our free-space optical communication. (b) Picture of the 1 km experiment field. (c) Picture of the FSO devices. Alice is fixed in the building and Bob is loaded on a radio-controlled electric vehicle. (d) Diffraction loss of our FSO system. Red: diffraction loss only; Blue: Diffraction loss with loss of optics.

The scheme of our free-space optical communication is shown in Fig. 1(a), and a pair of FSO devices are used at two nodes with separation between 1 and 4 km. Alice is fixed at the top floor in our building, and Bob is loaded on a radio-controlled electric vehicle (RCEV) so that it can move around to change the FSO link distance for the experiments, as shown in Fig. 1(b). At both ends, our FSO devices feature identical optical aperture and mechanical design, as shown in Fig. 1(c). Each FSO device consists an optical transceiver module, an APT unit and its control electronics, and both the transceiver module and the control electronics are sealed in a box for outdoor operation. Both FSO nodes are SMF-coupled, so that bidirectional optical communication can be achieved by the direct use of commercial optical transceiver modules. We choose the optical aperture of 90 mm in diameter and study the diffraction loss of the FSO link as a function of link distance. The result is shown in Fig. 1(d), and less than 12.7 dB low diffraction can be calculated within a link distance of 10 km, considering the internal transmission of our optics.

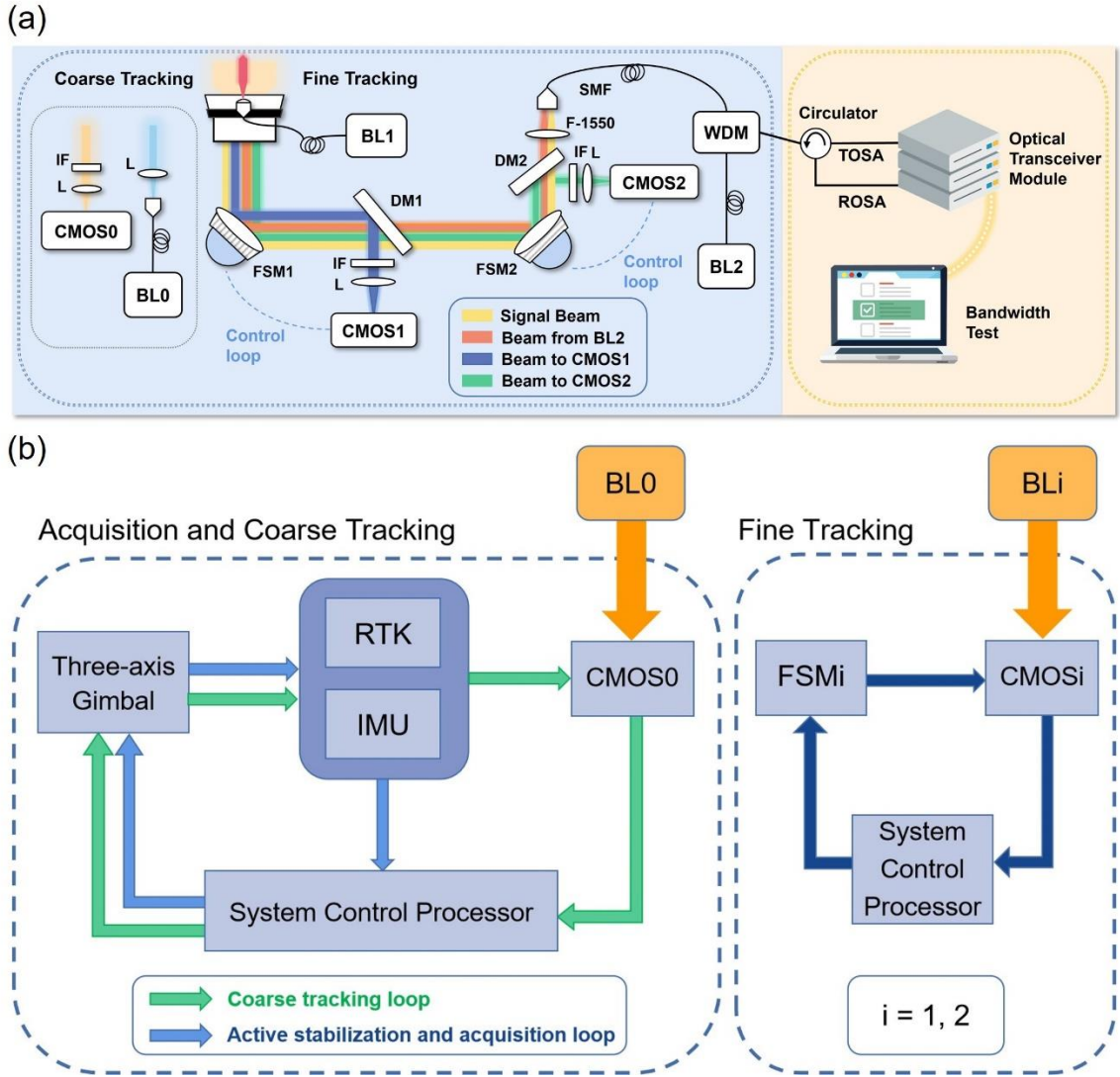

**FIG. 2.** (a) The design of an FSO device. L, lens; IF, interference filter; DM, dichroic mirror; WDM, wavelength division multiplexer; TOSA, transmitter optical sub-assembly module; ROSA, receiver optical sub-assembly module. CMOS, complementary metal oxide semiconductor; BL, beacon laser; FSM, fast steering mirror system. (i = 1, 2) (b) The operation schematic of the APT system for acquisition and coarse tracking (left), fine tracking (right).

Fig. 2(a) shows the design of an FSO device. The blue part in the left shows the schematic of the optical part in an APT system, which is mounted on the 3-axis gimbal stage for coarse tracking and active stabilization. The optical antenna is in a reflective Cassegrain design with 90 mm optical diameter and scale factor of 10:1, which is followed by the fine tracking optics. Three beacon laser (BL) diodes are used for the three-stage optical tracking, where BL0 (940 nm) and BL1 (638 nm for Alice, 660 nm for Bob)/BL2 (808 nm for Alice, 852 nm for Bob) are for the coarse tracking and first/second-stage fine tracking, respectively. Three complementary metal oxide semiconductor (CMOS) sensors, CMOS0 and CMOS1/CMOS2, are used to generate the error signals for coarse tracking and first/second-stage fine tracking, respectively. The divergent angles of the three BLs are designed to be overlapped with each other, so that APT process can be achieved stage by stage. While the beacon beam from BL0 and CMOS0 are placed off axis from the signal beam, the beacon beam from BL1 and BL2 are concentric to the signal beam. The beacon beam from BL2 is from the same fiber for FSO signal, and a wavelength division multiplexer (WDM) filter is used to separate this beacon laser and communication signal. This fiber is connected to the optical transceiver module, with an optical circulator in between, for the separation of the up/down link optical signals.

With Bob moving on the RCEV, the FSO link can be established within 10 minutes using our APT systems once settled. The operation schematic of the APT system is shown in Fig. 2(b), and its left part shows the mechanism of the active stabilization, initial acquisition and coarse tracking processes using a three-axis gimbal in the APT system. The active stabilization is achieved using the rotational angular acceleration signal from an inertial measurement unit (IMU). This active stabilization isolates the APT system from the mechanical vibration during the FSO and ensures the link stability. The initial acquisition is performed by feeding the relative angle between Alice and Bob to the three-axis gimbal. Such relative angle is calculated by sharing the accurate satellite positioning signals from real-time kinematic modules (RTK) mounted in the APT systems at Alice and Bob. Then the CMOS0 can detect the coarse tracking beam from BL0 in the opposite communication node for the error signal generation and takes over the control of the three-axis gimbal. Then the APT systems are ready for the fine optical tracking. The CMOS1 detects the fine tracking beam from BL1 and feedback the error signal into FSM1, and a second-stage fine tracking is performed using CMOS2 and FSM2 in a similar mechanism to further increase the tracking accuracy on top of the first-stage fine tracking loop, as shown in the right part of the Fig. 2(b). The detail parameters of the APT system are shown in Table I.

**Table I. Performance of the APT system**

| Components | | Value |
|---|---|---|
| Coarse tracking mechanism | Type | 3-axis motorized gimbal stage |
| | Tracking range | Azimuth: ± 90 ° Pitch: ± 60 ° (With Roll fixed) |
| Coarse tracking sensor | Type | CMOS |
| | FOV | 0.04 rad * 0.04 rad |
| | Size & Frame rate | 288 * 288 pixels & 1 kHz |
| Beacon laser0 | Power | 1 W |
| | Wavelength | 940 nm |
| | Divergence | 35 mrad |
| Fine tracking mechanism | Type | FSM |
| | Range | ±212 μrad |
| Fine tracking sensor | Type | CMOS |
| | FOV | 13 mrad * 10 mrad |
| | Size & Frame rate | 288 * 288 pixels & 1 kHz |
| Beacon laser1 | Power | 5 mW |
| | Wavelength | 638 nm & 660 nm |
| | Divergence | 6 mrad |
| Beacon laser2 | Power | 5 mW |
| | Wavelength | 808 nm & 852 nm |
| | Divergence | 6 mrad |

We first measure the performance of the APT system in a 1 km FSO link. Fig. 3(a) shows a coarse tracking error in a 120 s measurement, and the average error is 24 μrad, with standard deviation of 34.6 and 20.9 μrad for pitch and azimuth direction, respectively. Then we conduct a measurement of fine tracking performance, with result shown in Fig. 3(b). The fine tracking control is turned off in the first 30 s, and activated in the following 60 s. The result shows that the average tracking error is reduced rapidly once the fine tracking is tuned on, from 24 μrad to 3 μrad, with standard deviation of 2.9 and 3.9 μrad for pitch and azimuth direction, respectively. We measure the optical transmission loss using this FSO link, with results shown in Fig. 4(a). An average link loss is measured to be 29.3 dB with only first-stage fine tracking turned on (blue curve). The red curve shows the link loss with the second-stage fine tracking turned on, and it reaches 13.7 dB, with standard deviation of 1.4 dB.

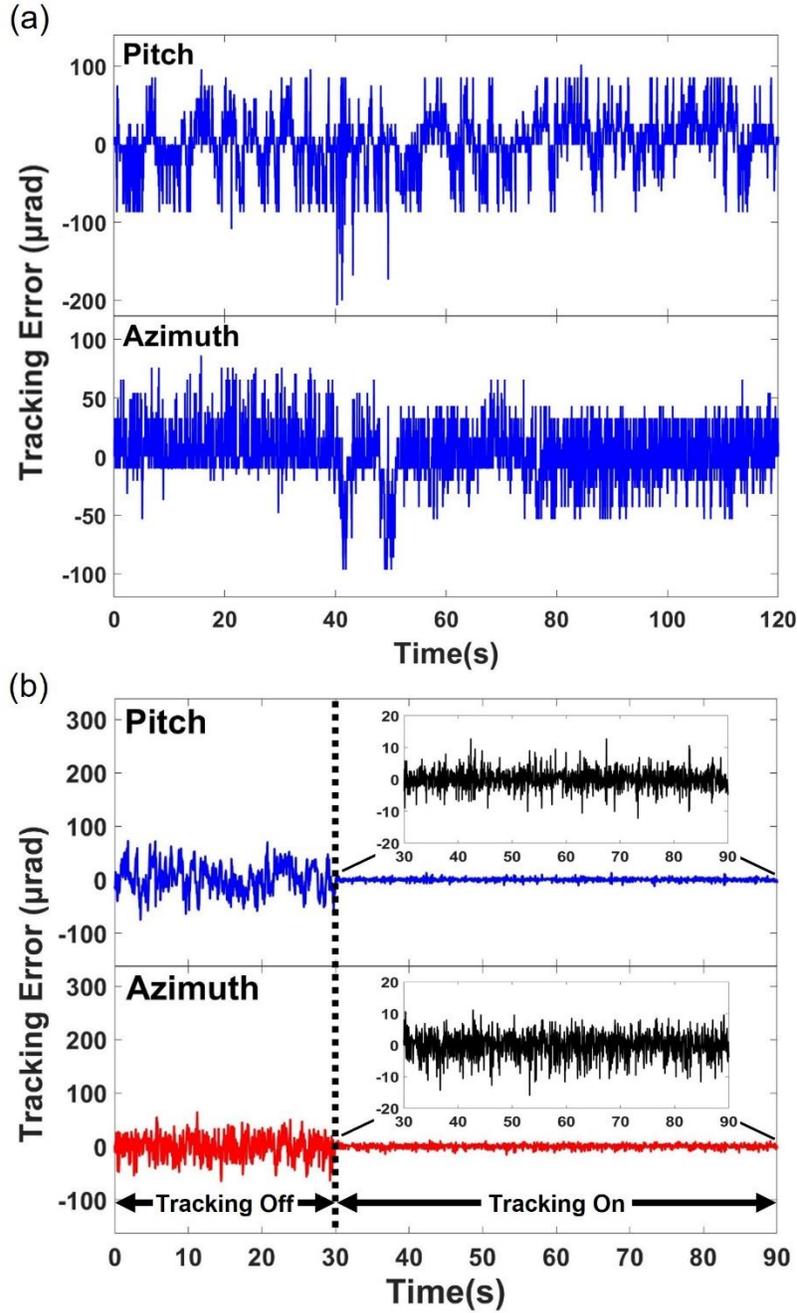

**FIG. 3.** Performance of the APT system measured with 1 km separation. (a) The coarse tracking error. (b) The fine tracking error.

Then we use a pair of standard commercial fiber optical transceiver modules for the FSO measurement. Both modules are rated at 10 Gbps, where the transmitter optical sub-assembly (TOSA) and receiver optical sub-assembly (ROSA) modules are both SMF-coupled in each module, as shown in Fig. 4(b). We first test the performance of these modules, by direct fiber connections between Alice and Bob in their TOSA and ROSA ports, respectively. Using adjustable optical attenuators, identical losses can be applied for both links. The communication bandwidth is measured following the transmission control protocol (TCP), via the software "open-source tool iperf3" (ESnet and Lawrence Berkeley National Laboratory) .[15] As shown in Fig. 4(c), the average communication bandwidth is measured at 9.27 Gbps during a 100 s test, and a maximum link loss exceeds 24.1 dB to reach such full bandwidth operation using these modules. Through the 1 km FSO link, the average communication bandwidth is measured to be 9.16 Gbps in a 100 s test, with a standard deviation of 0.24 Gbps. Hence almost identical communication bandwidth is achieved in FSO.

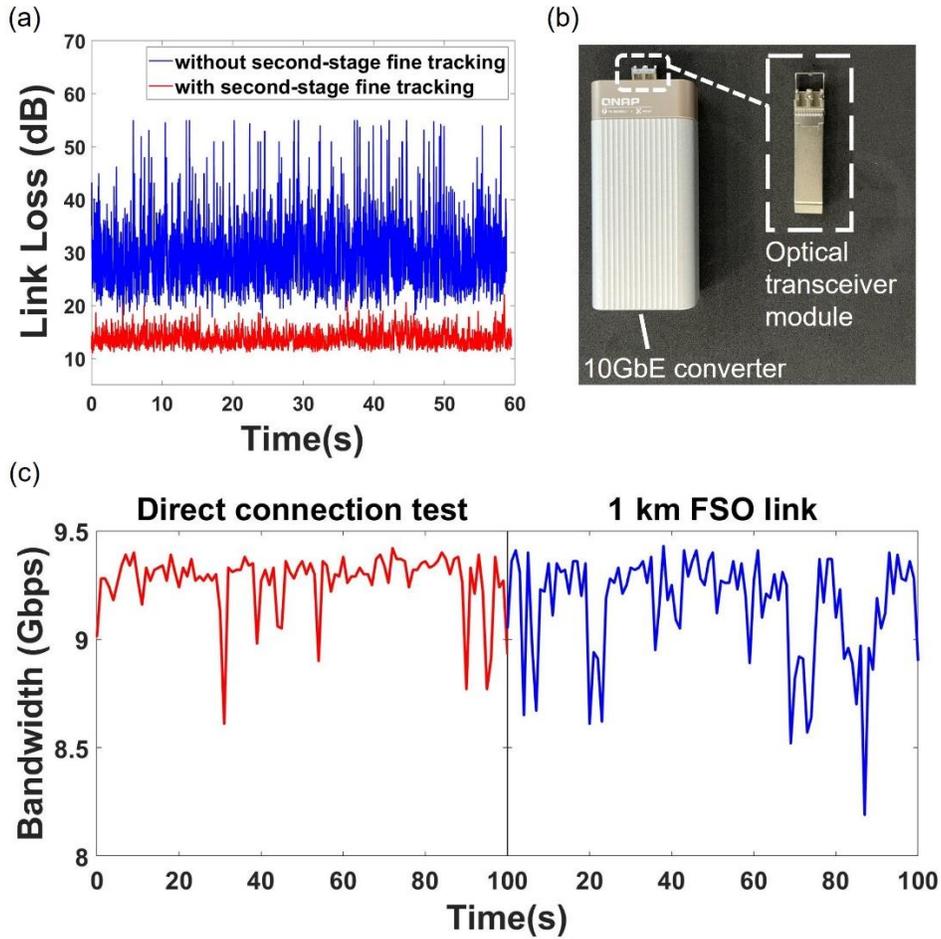

**FIG. 4.** (a) Link loss for 1 km FSO. (b) Picture of the optical transceiver modules. (c) Communication bandwidth measurement for the module test and FSO. Red: direct connection test; Blue: 1 km FSO test.

We further measure the APT performance in a 4 km FSO link, where the average link loss is measured to be 18 dB with standard deviation of 2.8 dB, as shown in Fig. 5. Compared with the 1 km link, the average link loss increases 4.3 dB. The maximum loss is 27.8 dB, which exceeds the tolerable maximum loss of the transceiver module, hence stable FSO link can hardly be achieved. Our experiment was conducted during the foggy winter in Nanjing with the air visibility generally around 4-5 km, which corresponds to absorption loss of ~4 dB in 4 km distance.[16] With better air quality or more sensitive optical transceiver module, we believe our FSO system can achieve optical communication in the 4 km or longer distances.

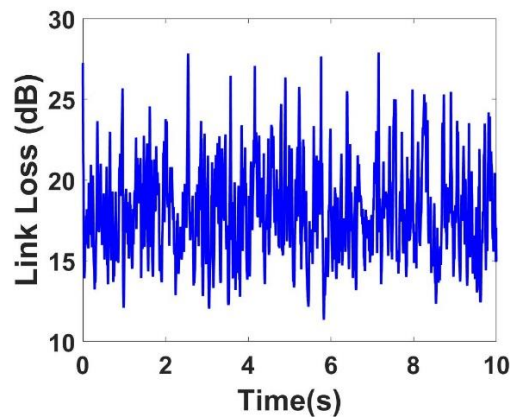

**FIG. 5.** Link loss for 4 km FSO.

In this work, we demonstrate 1 km FSO with communication bandwidth of 9.16 Gbps, using a pair of commercial fiber optical transceiver modules without optical amplification. This result shows that an FSO network may be achieved using standard fiber communication devices and network topology. Compact FSO devices are developed, with weight of only 9.5 kg. It is fast field deployable, and can be installed on a RECV for quick link change and FSO establish within 10 minutes. We test our FSO link up to 4 km, and a maximum loss of 27.8 dB is measured. At such loss level, FSO with the same high bandwidth can be expected in a longer range with the use of EDFAs. In this experiment, we only test the FSO with both nodes fixed. However, our FSO system can also work when the RCEV is moving slowly, and hence mobile FSO link can be expected in the following study. Beside the classical FSO application, our compact APT systems can also be used to establish quantum links for quantum communication applications, including quantum secure direct communication,[17] and quantum information transmission,[18-20] etc.

**Acknowledgments.** This work was supported by the National Key R&D Program of China (No. 2019YFA0705000), Leading-edge technology Program of Jiangsu Natural Science Foundation (No. BK20192001), National Natural Science Foundation of China (51890861, 11690033, 62293523), Zhangjiang Laboratory (ZJSP21A001), Key R&D Program of Guangdong Province (2018B030329001), the National Postdoctoral Program for Innovative Talents (BX2021122), China Postdoctoral Science Foundation (No. 2022M711570), the Fundamental Research Funds for the Central Universities (2022300158), and Jiangsu Funding Program for Excellent Postdoctoral Talent.

**Conflict of Interest.** The authors have no conflicts to disclose.

**Data availability statement.** The data that support the findings of this study are available from the corresponding author on reasonable request.